\documentclass[aps,prd,twocolumn,groupedaddress,showpacs,showkeys,draft]{revtex4-1}
\usepackage{graphicx}
\usepackage{dcolumn}
\usepackage{bm}
\usepackage{amssymb,amsmath,latexsym,amsfonts}
\begin{document}


\title{Superfluidity as a Tool for Constraining the Energy--Momentum Relation}

 \author{A. Camacho}
 \email{acq@xanum.uam.mx} \affiliation{Departamento de F\'{\i}sica,
 Universidad Aut\'onoma Metropolitana--Iztapalapa\\
 Apartado Postal 55--534, C.P. 09340, M\'exico, D.F., M\'exico.}

 \author{S. Guti\'errez}
 \email{sergiogs@xanum.uam.mx} \affiliation{Departamento de F\'{\i}sica,
 Universidad Aut\'onoma Metropolitana--Iztapalapa\\
 Apartado Postal 55--534, C.P. 09340, M\'exico, D.F., M\'exico.}


\date{\today}

\begin{abstract}
The present work explores the possibilities that superfluidity could
offer in the context of quantum gravity phenomenology, at least in
the realm of deformed dispersion relations. The experimental
proposal involves a Bose--condensed sodium gas trapped by an
isotropic harmonic oscillator potential. A deformed dispersion
relation for the particles of the system is considered and the
consequences of this assumption upon the possible region of
superfluidity of this system is analyzed. It will be shown that in
this sense the effects of quantum gravity could be detected
resorting to experiments of superfluidity in Bose--Einstein
condensates. Finally, using the current experimental results in this
direction an upper bound for the corresponding phenomenological
parameters will be also obtained.
\end{abstract}

\maketitle

\section{Introduction}

Quantum gravity phenomenology \cite{Giov5,
Giov15,FischbachTalmadge99} emerges as an answer of the community to
the mathematical and physical difficulties plaguing all theoretical
models behind a quantum theory of gravity \cite{kop04, beck1}. These
efforts embody new physical effects, for instance, deformed versions
of the dispersion relation, deviations from the $1/r$--potential and
violations of the equivalence principle. At this point it is
noteworthy to comment that these cases do not exhaust the extant
possibilities.

In the quest for a solution to this long--standing puzzle in modern
physics we have some efforts which entail, unavoidably, the
breakdown of Lorentz symmetry \cite{Amelino1, Amelino2, Amelino3}.
Lorentz symmetry is a the bedrock of modern physics and, therefore,
it has been subjected to some of the highest precision tests in
Physics \cite{Robertson, Claus1, Schiller1}. The current
experimental results show no evidence of a violation of this
symmetry, nevertheless this last fact does not discard it and, in
consequence, further work is required. At this point it has to be
clearly stated that the phrase violation of Lorentz symmetry has
several meanings, i.e., it embodies several characteristics. For
instance, Local Lorentz Invariance, or Local Position Invariance
\cite{Will1}. For us it will mean a modification of the dispersion
relation. Let us state this phrase in a mathematical language.
 As mentioned above several quantum--gravity models predict a
modified dispersion relation \cite{Amelino1, Amelino2, Amelino3},
the one can be characterized, phenomenologically, through
corrections hinging upon Planck's energy, $E_p$

\begin{equation}
E^2 = p^2c^2\Bigl[1 - \alpha\Bigl(E/E_p\Bigr)^n \Bigr] + (mc^2)^2.
\label{Disprel2}
\end{equation}

Here $\alpha$ is a coefficient, whose precise value depends upon the
considered quantum--gravity model, while $n$, the lowest power in
Planck's length leading to a non--vanishing contribution, is also
model dependent.

The quest in this direction has already considered interferometry as
a tool \cite{Amelino4, Camacho1, Camacho2, Camacho3, Elias}.
Clearly, all these previous works in quantum gravity phenomenology
involve different fields of physics \cite{Amelino77}, i.e., the
search includes now many areas of modern physics. One of these
topics is condensed matter physics. Indeed, the use of cold atoms,
either bosonic or fermionic, is a point already considered
\cite{Coll1, Coll2}. In particular the possibility of constraining
the energy--momentum relation resorting to cold atoms has already
shown us that this kind of systems could open up new landscapes in
the context of gravitational physics \cite{Giovanni34}. Of course,
this last topic in the context of phenomenology of quantum gravity
leads us to ask if there are additional low--temperature effects
that could be used as trackers for new effects. This question
implies the quest for the detection of these kind of effects in the
realm of condensed matter physics, i.e., a broadening of the current
attempts. Clearly, another low--temperature effect is the phenomenon
of superfluidity \cite{Nozieres} and, therefore, we wonder if this
case could offer a new window for our search. In the present work we
explore this situation, namely, the possibilities that superfluidity
has to offer in the context of quantum gravity phenomenology.

Having stated our goal we must discuss, though briefly, the physics
behind the emergence of viscosity in a flow. Concerning the
phenomenon of superfluidity the first experimental results can be
found in the work of Kamerling Onnes of 1911 in which he detected
that if cooled below $2.2 ~^{\circ}K$ $He$ did not contract but
rather expand \cite{Onnes}. The current work has been able to
provide a coherent picture to the subjacent Physics
\cite{Khalatnikov, Griffin}. Bose--Einstein Condensation (BEC) is
also connected to the presence of very low temperatures. Fritz
London \cite{London} put forward the idea of a connection between
these two effects asserting that the transition from $He~I$ (the
high temperature phase of liquid helium) and $He~II$ (the low
temperature phase) should be considered an example of a BEC. Taking
into account London's idea and, joining it to the previous work
BEC--quantum gravity phenomenology, once again, we are confronted
with the question about a possible use of superfluidity as a tool in
our quest.

The concept of elementary excitations in the realm of superfluidity
was first introduced by Landau \cite{Landau1} (within the two--fluid
model proposed by Tisza \cite{Tisza}) as a core feature in the
description of the behavior of $He~II$. Landau asserted
\cite{Landau2} that the normal fluid (the non--superfluid component)
could be regarded as a dilute gas whose components are
weakly--interacting elementary excitations which move in a
background defined by the superfluid component. Along these ideas
the phenomenon of superfluidity appears when the velocity of the
corresponding flow lies below a certain threshold value given by

\begin{equation}
v_{(crit)}= \min\Bigl(\frac{\epsilon(p)}{p}\Bigr). \label{Landau}
\end{equation}

In this last expression $\epsilon(p)$ denotes the energy of an
elementary excitation and $p$ the corresponding momentum.

In the experimental realm the quest for this critical velocity has
been carried out in a sodium--BEC, and the results show a possible
velocity threshold located around the value of $1.6~mm/s$
\cite{Raman}. Sodium is a system that can be condensed and the speed
of sound in it has already been detected \cite{Andrews, Andrews1}.

At this point we may now state clearly the ideas contained in the
present work and why they have been chosen. The main purpose is to
obtain a prediction for the critical velocity for a BEC in which the
relation energy--momentum of the particles of the gas has been
deformed along some possibilities contained in several quantum
gravity models. The system is a sodium gas trapped by an isotropic
harmonic oscillator potential. The reason for this particular choice
stems from the fact that several of its properties have already been
detected and measured, for instance, evidence for a critical
velocity \cite{Raman} and speed of sound \cite{Andrews, Andrews1}.
In other words, our proposal is the following one: use a
BEC-condensed sodium gas and measure the region in which
superfluidity is present. Compare the size of this region against
the theoretical prediction here obtained and deduce the
corresponding bound for our parameter containing the breakdown of
Lorentz symmetry. Up to now, superfluidity has not been considered a
relevant element in quantum gravity phenomenology and the present
work shall be considered as an analysis of the perspectives that
this topic could offer.

\section{Mathematical Model}

Even the simplest mathematical model for a BEC (the Gross--Pitaevski
equation) trapped by a harmonic oscillator potential has no
analytical solution, yet \cite{Lieb1}. This last fact implies that
in this topic we must resort to approximation methods, etc. In our
case, the theoretical model will describe the features of the BEC
resorting to an approximation method in which the presence of
interactions among the particles produces a change in the frequency
of the trap rendering a smaller value than the one provided by the
trap, for the case of repulsive interactions. This assumption will
allows us to calculate the energy of the ground state and of the
thermal cloud. The energy and momentum of the elementary
excitations, according to Bogoliubov ideas, are a function of the
energy of the excited particles, and this parameter can be computed
from our assumptions. Finally, the critical velocity is deduced as a
function of our phenomenological variables and compared against the
measurement readouts. From this comparison an upper bound for our
model values will appear.

From a fundamental point of view our mathematical model can be
defined by an $N$--particle Hamiltonian the one in the formalism of
second quantization is \cite{Ueda}

\begin{eqnarray}
\hat{H}= \int d\vec{r}\Bigl[-\hat{\psi}^{\dagger}(\vec{r}, t)
\frac{\hbar^2}{2m}\nabla^2\hat{\psi}(\vec{r}, t)\nonumber\\
+ V(\vec{r})\hat{\psi}^{\dagger}(\vec{r}, t)\hat{\psi}(\vec{r}, t)\nonumber\\
+\frac{U_0}{2}\hat{\psi}^{\dagger}(\vec{r},
t)\hat{\psi}^{\dagger}(\vec{r}, t) \hat{\psi}(\vec{r},
t)\hat{\psi}(\vec{r}, t)\Bigr]. \label{Hamilton1}
\end{eqnarray}

In this Hamiltonian $\hat{\psi}^{\dagger}(\vec{r}, t)$ and
$\hat{\psi}(\vec{r}, t)$ represent
 bosonic creation and annihilation operators, respectively.
 It is valid only at low energies
and momenta and implies that the interaction among the particles is,
as usual, codified by the scattering length parameter $a$, i.e.,
$U_0=\frac{4\pi a\hbar^2}{m}$. The corresponding trapping potential
($V(\vec{r})$) is an isotropic harmonic oscillator whose frequency
reads $\omega$. Moreover, there are $N$ particles in the gas, each
of them with mass $m$, the volume occupied by the system is $V$.

The  mathematical assumptions in the present model read:

(i) Only two states are populated, namely, ground and the first
excited state. We may provide to this assumption a realistic
physical meaning recalling that for a bosonic system, with chemical
potential $\mu$ and energy levels of single--particle $\epsilon$,
the occupation number in thermal equilibrium is given by
\cite{Pethick} ($\beta=1/(\kappa T$))

\begin{eqnarray}
<n_{(\epsilon)}>=\frac{1}{e^{(\epsilon-\mu)\beta}-1}.\label{Occunumber}
\end{eqnarray}

Clearly, it is a monotonic decreasing function of $\epsilon$, and
this feature justifies the present assumption.

(ii) The Hartree approximation will be employed for the mathematical
description of the two occupied states. In other words, the ground
state of the interacting system is deduced by a
Ginzburg--Pitaevski--Gross energy functional \cite{Gross1}, and it
entails that the ground state wavefunction corresponds to the case
of a harmonic oscillator situation but the frequency is modified due
to the fact that the system has a non--vanishing scattering length
\cite{Baym}, such that the fundamental length parameter reads.

\begin{equation}
R= \Bigl(\frac{2}{\pi}\Bigr)^{1/10}\Bigl(\frac{Na}{l}\Bigr)^{1/5}l.
\label{Radius1}
\end{equation}

In this last expression $l$ is the radius related to the trap given
by the isotropic harmonic oscillator of the trap

\begin{equation}
l= \sqrt{\frac{\hbar}{m\omega}}. \label{Radius2}
\end{equation}

Of course, we end up with an effective frequency

\begin{equation}
\tilde{\omega}= \frac{\hbar}{mR^2}. \label{Freq1}
\end{equation}

The experimental conditions entail $R>l$ \cite{Raman} and,
therefore, $\tilde{\omega}<\omega$.

The order parameter of those  particles in the ground state is

\begin{eqnarray}
\psi_{(0)}(\vec{r})=\sqrt{\frac{N_{(0)}}
{(R\sqrt{\pi})^3}}\exp{\Bigl[-\frac{r^2}{2R^2}\Bigr]}.\label{Groustate}
\end{eqnarray}

Here $N_{(0)}$ denotes the number of particles in the lowest energy
state. The presence of a non--vanishing scattering length implies
that in the ground state not all the particles can have
zero--momentum, the reason for this lies in the fact that the
two--body interaction mixes in components with atoms in other states
\cite{Pethick} and

\begin{eqnarray}
N_{(0)}=N\Bigl[1-\frac{8}{3}\sqrt{\frac{Na^3}{\pi
V}}\Bigr].\label{Depletion}
\end{eqnarray}

Clearly,

\begin{eqnarray}
N_{(0)}= \int\bigl(\psi_{(0)}(\vec{r})\bigr)^2d^3r,\label{Norm1}
\end{eqnarray}

\begin{eqnarray}
V=\frac{4\pi}{3}R^3.\label{Volume1}
\end{eqnarray}

Concerning the thermal cloud, the core of this part is also
comprised by the Hartree approximation. We assume that all excited
particles are in the same state and, since the temperature is very
low, it corresponds to the first excited state of a particle trapped
by a harmonic oscillator with a frequency given by (\ref{Freq1}). If
$\Psi_{(1)}$ denotes the wave function of the thermal cloud,
$\phi_{(1)}$ the wavefunction of the first excited state of a single
particle in our effective trap, and $N_{(e)}$ the number of
particles in the cloud then

\begin{eqnarray}
\Psi_{(1)}(\vec{r})=\sqrt{N_{(e)}}\phi_{(1)}(\vec{r}).\label{Hartree1}
\end{eqnarray}

From the symmetry of the system we may conclude that an excited
particle can have vanishing momentum along the $x$ and $y$ axes but
a non--zero one in the $z$--direction, or vanishing momentum along
the $z$ and $y$ axes but non--zero one in the $x$--direction, or,
finally, vanishing momentum along the $x$ and $z$ axes and larger
than zero along $y$--direction. The symmetry of our trap and of the
scattering length entail that one third of the excited particles
will have non--vanishing momentum along the $x$--axis, one third
along the $y$--axis, and the remaining third along the
$z$--direction. Mathematically this corresponds to the following
expressions

\begin{eqnarray}
\psi^{(i)}_{(1)}(\vec{r})=\frac{8}{\sqrt{27\pi}}\sqrt{\frac{N}{V}\sqrt{\frac{Na^3}{\pi
V}}}\frac{x^{(i)}}{R}\exp{\Bigl[-\frac{r^2}{2R^2}\Bigr]}.\label{Excstate}
\end{eqnarray}

Here $x^{(1)}=x$, $x^{(2)}=y$, and $x^{(3)}=z$.

Of course, (\ref{Excstate}) must be related to the total number of
particles in excited states
($N_{(e)}=\frac{8}{3}N\sqrt{\frac{Na^3}{\pi V}}$), a condition that
becomes \cite{Pethick}

\begin{eqnarray}
N_{(e)}=\int\Bigl[\sum_{i=1}^{3}
\bigl(\psi^{(i)}_{(1)}(\vec{r})\bigr)^2\Bigr]d^3r.\label{Norm2}
\end{eqnarray}

The three expressions in (\ref{Excstate}) will be used for the
definition of the wavefunction ($\Psi_{(1)}(\vec{r})$) of the
thermal cloud, i.e.,

\begin{eqnarray}
\Psi_{(1)}(\vec{r})= \psi^{(x)}_{(1)}(\vec{r})+
\psi^{(y)}_{(1)}(\vec{r})+\psi^{(z)}_{(1)}(\vec{r}).\label{Excstate1}
\end{eqnarray}

\section{Elementary Excitations and Deformed Dispersion Relations}

Having stated our assumptions we proceed to compute the critical
velocity \cite{Nozieres}. The deduction of the energy of an
elementary excitation and of its corresponding momentum requires the
knowledge of the energy of a single--particle in the first excited
state \cite{Ueda}. The thermal cloud contains particles in the first
excited state of an isotropic harmonic oscillator whose frequency is
(\ref{Freq1}) therefore the energy of an excited particle is given
by this assumption and easily calculated as a function of the
effective frequency of our variational procedure

\begin{eqnarray}
\tilde{\epsilon}^{(0)}=
\frac{5}{2}\hbar\tilde{\omega}.\label{excene1}
\end{eqnarray}

This is the case in which no deformed dispersion relation has been
considered. Introducing this quantum gravity parameter we have

\begin{eqnarray}
\tilde{\epsilon}= \frac{5}{2}\hbar\tilde{\omega} +\alpha
p^n.\label{excene01}
\end{eqnarray}

In this last expresion $\alpha$ and $n$ are parameters stemming from
the considered quantum gravity model \cite{Amelino77}.

According to the ideas of Bogoliubov \cite{Ueda, Bogoliubov} the
energy of an elementary excitation, here denoted by $\epsilon$, is a
function of the energy of the excited particles of the BEC, namely,

\begin{eqnarray}
\epsilon=\sum\sqrt{(\tilde{\epsilon})^2+\frac{2NU_{(0)}}{V}\tilde{\epsilon}}.\label{excene2}
\end{eqnarray}

The contribution to the energy of all the elementary excitations
turns out to be \cite{Ueda, Bogoliubov}

\begin{eqnarray}
\tilde{E}=\sum\sqrt{(\tilde{\epsilon})^2+\frac{2NU_{(0)}}{V}\tilde{\epsilon}}<\tilde{n}_{\epsilon}>.\label{Excene2}
\end{eqnarray}

We have defined  $<\tilde{n}_{\epsilon}>$ as the occupation number
of the elementary excitations with energy ${\epsilon}$. The relation
between the occupation numbers of particles and elementary
excitations is \cite{Ueda}

\begin{eqnarray}
<\tilde{n}_{\epsilon}>=
\frac{<n_{\epsilon}>}{1+<n_{\epsilon}>}.\label{Occunumber3}
\end{eqnarray}

Since our model has to be consistent with the present experimental
technology at this point we resort to the laboratory values related
to the detection of a critical velocity in a sodium condensed gas
\cite{Raman} in which the occupation number of the particles in the
first excited state fulfills the condition $N_{(e)}\sim 10^2>1$.
Therefore, $<\tilde{n}_{\epsilon_{(1)}}>=1$. In addition,
$<\tilde{n}_{\epsilon_{(i)}}>=0, ~~\forall i>1$. Indeed, we have
considered that the thermal cloud is comprised by particles which
occupy only the first excited state, in other words,
$<n_{\epsilon_{(i)}}>=0,~~\forall i>1$. Introducing this condition
into (\ref{Occunumber3}) leads us to the aforementioned result for
the occupation number of the elementary excitations.

The order of magnitude of this deformed dispersion relation has to
be very small, otherwise, it would have already been detected. This
means that we must expect the fulfillment of

\begin{eqnarray}
{\epsilon}^{(0)}>>\alpha p^n.\label{Unequ1}
\end{eqnarray}

We now cast (\ref{excene2}) in a different form, and for this we
resort to the effective volume $V=4\pi R^3/3$, use (\ref{Radius1}),
(\ref{Radius2}), and (\ref{Freq1}) and keep only terms linear in
$\alpha$. The final result reads

\begin{eqnarray}
\epsilon=\Bigl(\frac{4\pi}{3}\Bigr)^{1/3}\frac{\hbar^2}{mV^{2/3}}\sqrt{\frac{25}{4}\Bigl(\frac{4\pi}{3}\Bigr)^{2/3}+
\frac{20\pi Na}{V^{1/3}}}
\nonumber\\
+\frac{\tilde{\alpha}}{2}\sqrt{\frac{4\pi
Na}{5V^{1/3}}}\frac{1}{V^{n/3}},\label{excene3}
\end{eqnarray}

\begin{eqnarray}
\tilde{\alpha}=\Bigl(\frac{4\pi}{3}\Bigr)^{2/3}\Bigl[\sqrt{24\pi}\Bigl(\frac{4\pi}{3}\Bigr)^{1/3}\hbar\Bigr]^n\alpha
.\label{excene13}
\end{eqnarray}

According to Landau \cite{Landau2}, in order to find the critical
velocity we must now deduce the momentum of this elementary
excitation. These physical variables, which define the normal
component of the fluid fluid, can be regarded as a bosonic gas whose
components are weakly--interacting and moving in a region in which a
constant potential exists, and this potential is defined by a mean
field approach \cite{Ueda}. According to this interpretation we may
rewrite (\ref{excene3}) in the same form as in the case in which our
BEC is a homogeneous one \cite{Ueda}. In other words, take
(\ref{excene3}) impose the condition $\alpha =0$ and compare the
result against

\begin{eqnarray}
\epsilon=\frac{\hbar^2k}{2m}\sqrt{k^2+ \frac{16\pi
Na}{V}}.\label{excene4}
\end{eqnarray}

This last argument allows us to deduce the wavenumber related to our
elementary excitation and, therefore, its momentum.

\begin{eqnarray}
k=\Bigl(\frac{4\pi}{3}\Bigr)^{1/3}\sqrt{5}\frac{1}{V^{1/3}}
,\label{elemwavenum}
\end{eqnarray}

\begin{eqnarray}
p=\Bigl(\frac{4\pi}{3}\Bigr)^{1/3}\sqrt{5}\frac{\hbar}{V^{1/3}}
.\label{elemmom}
\end{eqnarray}

Resorting to Landau criterion (\ref{Landau}) we obtain that the
critical velocity is given by

\begin{eqnarray}
v_{(crit)}=\frac{1}{\sqrt{5}}\frac{\hbar}{mV^{1/3}}\sqrt{\frac{25}{4}\Bigl(\frac{4\pi}{3}\Bigr)^{2/3}+
\frac{20\pi Na}{V^{1/3}}}\nonumber\\
+\frac{1}{\sqrt{20}}\Bigl(\frac{4\pi}{3}\Bigr)^{1/3}\nonumber\\
\times\Bigl[\sqrt{24\pi}\Bigl(\frac{4\pi}{3}\Bigr)^{1/3}\Bigr]^n
\sqrt{\frac{4\pi
Na}{5V^{1/3}}}\Bigl(\frac{\hbar}{V^{1/3}}\Bigr)^{n-1}\alpha
.\label{crit2}
\end{eqnarray}

\section{Critical Velocity}

At this point we proceed to check our model, and in order to do this
we consider the case in which $\alpha=0$. The experimental
parameters \cite{Raman} to be used are: (i) a critical speed of
$v^{(e)}_{(crit)}= 1.6~mm/s$; (ii) the number of particles in this
experiment has a minimum of $N=3\times 10^6$ and a maximum of
$N=12\times 10^6$, and for the evaluation of our expression we will
take the arithmetic average, i.e., $N=7.5\times 10^6$; (iii) the
effective volume is that of an ellipsoid whose axes are
$l_1=45\times 10^{-6}m$ and $l_1=150\times 10^{-6}m$ such that
$V=\frac{4\pi}{3}l^2_1l_2$, and, finally, (iv) a scattering length
$a=2.75\times 10^{-9}m$.

Introducing these values into (\ref{crit2}) (setting $\alpha=0$)
implies

\begin{eqnarray}
v^{(m)}=1.95~mm/s.\label{crit3}
\end{eqnarray}

The reported critical speed is \cite{Raman}

\begin{eqnarray}
v^{(e)}= 1.6~mm/s
\end{eqnarray}

The ensuing error is less that 18 percent

\begin{eqnarray}
\vert v^{(e)}-v^{(m)}\vert/\Bigl(v^{(m)}\Bigr)=0.179.\label{error1}
\end{eqnarray}

In other words, our model provides a very good description of the
experiment and, hence, the analysis of the deformed case within the
present framework seems a reasonable assumption.

\section{Conclusions}

In the present work the analysis of the options that superfluidity
could offer in the context of quantum gravity phenomenology has been
done. The model has been a Bose--condensed sodium gas trapped by an
isotropic harmonic oscillator in which the energy--momentum relation
for the particles has been deformed along the proposals emerging
from some quantum gravity models. Afterwards, along a perturbative
approach, the energy and momentum of the elementary excitations
generated by the particles in the thermal cloud have been
calculated. Finally, we introduce these last two physical parameters
in Landau criterion associated to superfluidity and find that the
breakdown of Lorentz symmetry, in the form of a deformed dispersion
relation, implies a modification of the region in which, for a
sodium BEC, superfluidity may exist. If $\alpha>0$, then the
aforementioned region grows (compared to the case in which
$\alpha=0$), whereas, if $\alpha <0$, then this region becomes
smaller. Indeed, the allowed superfluidity velocities are those
falling into the interval $(0, v_{(crit)})$. and, clearly, this
interval becomes larger for $\alpha>0$.

In relation with a bound for our phenomenological parameter
$\alpha$, a fleeting glimpse at (\ref{crit2}) tells us that we must
first choose a value for $n$. As an example we take $n=1$ (a
condition that implies that $\alpha$ has units of speed) and
consider the experimental values related to the evidence of a
critical velocity in a sodium condensed gas \cite{Raman}. Clearly a
choice has to be made in connection with the number of particles
since this physical variable changed from experiment to experiment;
for the sake of concreteness we consider the highest value, i.e.,
$N=12\times 10^6$, taking the lowest case ($N=3\times 10^6$) does
not modify the order of magnitude of the ensuing bound.

The measurement readouts \cite{Raman} are given up to units of
tenths of $mm/s$ and, in consequence, the smallest scale can be
considered as an approximation for the experimental error
\cite{Riveros}. In other words, if $\Delta v$ denotes the
experimental error of the measuring device, then the aforementioned
argument implies $\Delta v\sim 0.1~mm/s$. The experimental error has
to be equal or larger than the term containing the effects of the
breakdown of Lorentz symmetry, this phrase means for this situation

\begin{eqnarray}
14\times 10^{-5}~mm/s\geq\alpha. \label{error11}
\end{eqnarray}

Additional cases ($n=2$ does not imply new physics, it entails only
a redefinition of the inertial mass) can be analyzed in the same
manner. The present argument tells us that we may deduce an upper
bound for deformed dispersion relations associated to the structure
given by (\ref{Disprel2}), and this expression reads

\begin{eqnarray}
\Delta v\geq
\Bigl[\sqrt{24\pi}\Bigl(\frac{4\pi}{3}\Bigr)^{1/3}\Bigr]^n
\sqrt{\frac{4\pi
Na}{5V^{1/3}}}\Bigl(\frac{\hbar}{V^{1/3}}\Bigr)^{n-1}\alpha.
\label{error15}
\end{eqnarray}

Let us comment that in the present model the deformed dispersion
relation has been introduced only in the context of particles (see
(\ref{excene01})) but not in connection with the corresponding
elementary excitations also called quasi--particles, expression
(\ref{excene3})). Clearly, an additional possibility is the
introduction of the breakdown of Lorentz symmetry at the level of
the kinematics of the quasi--particles. This second option implies
the introduction of a second pair of phenomenological parameters,
since the quantum gravity modifications could be
particle--dependent. This case can be, without any further problem,
be considered in the present framework.

Summing up, we have shown that superfluidity in sodium--condensed
gases offers a new window in the realm of quantum gravity
phenomenology and that the present experimental results are enough
to deduce some rough bound for the involved parameters.

\begin{acknowledgements}
SG acknowledges CONACyT grant No. 324663.
\end{acknowledgements}

\end{document}